\documentclass[reprint,3p,twocolumn,sort&compress]{elsarticle}
\usepackage{amsfonts} 
\usepackage{amsmath}
\usepackage{amssymb}
\usepackage{graphicx}%\[  \]
\usepackage{subfigure}
\usepackage{color}
\journal{Journal of Atmospheric and Solar-Terrestrial Physics}
\begin{document}
\begin{frontmatter}
\title{Dynamical properties of   acoustic-gravity waves in the atmosphere}
\author[APM]{Animesh Roy}
\ead{aroyiitd@gmail.com}
\author[APM]{Subhrajit Roy}
\ead{suvo.math88@gmail.com}
\author[APM]{A. P. Misra\corref{cor1}}
\ead{apmisra@visva-bharati.ac.in; apmisra@gmail.com}
\cortext[cor1]{Corresponding author}
\address[APM]{Department of Mathematics, Siksha Bhavana, Visva-Bharati University, Santiniketan-731 235, India}
 %\pacs{52.27.Aj;  52.27.Ny; 52.35.Fp }
\begin{abstract}
We study  the dynamical behaviors of a system of five    coupled nonlinear equations that describes  the dynamics of acoustic-gravity waves in the atmosphere. A linear stability analysis together with the analysis of Lyapunov exponents spectra  are performed to show that the system can develop from ordered structures to   chaotic states.   Numerical simulation   of the system of equations reveals that  an interplay between the order and chaos indeed exists depending on whether the control parameter $s_1$, associated with the density scale height of acoustic-gravity waves, is below or above its critical value. 
\end{abstract}
\end{frontmatter}

%\textbf{Keywords:} acoustic gravity wave, stability, chaos, nonlinear dynamics.
\section{Introduction}\label{sec-intro}
%A brief introduction about the  acoustic gravity wave in the atmosphere is given by Potemra[3] in his thesis. He considered the atmosphere as an electrical transformer in which pressure is consider as the voltage of the electrical system. Therefore small change of pressure in the ground state generate the big change of pressure at the atmospheric level. Now the higher degree air compression due to gravity at the earth atmospheric level decreases  air density and pressure exponentially with altitude. In this scale the wavelengths are short at the wave propagation comparing with ground state, only pressure and inertial forces may be considered to define these disturbances as acoustic waves. Low-frequency disturbances with small amplitude wavelengths at the atmosphere are affected by the gravitational field and are called "acoustic-gravity waves".
The nonlinear dynamics of low-frequency finite amplitude acoustic-gravity waves has been studied by a number of authors because of their relevance in atmospheric disturbances \cite{stenflo1987,stenflo1991,stenflo1995,stenflo1996,Jovanovic2002,mendonca2015,kaladze2008}. The latter appear due to various meteorological conditions  including different  pressure and density gradients, as well as the presence of shear flows \cite{Jovanovic2002}. It has been shown that the nonlinear acoustic-gravity waves can appear in the forms of localized solitary   vortices \cite{stenflo1987,Jovanovic2002}, ordered structures \cite{park2016}, as well as chaos \cite{banerjee2001} and turbulence \cite{shaikh2008}.  
\par   
 In a paper \cite{stenflo1996}, Stenflo  deduced a  system of five coupled equations that describes the essential features of low-frequency atmospheric disturbances. His starting point was the most commonly used model equations for two-dimensional acoustic-gravity waves  of the form \cite{stenflo1995}
  \begin{equation}
D_t\left(\nabla^2\psi - \frac{1}{4H^2}\psi\right) = -\partial_x \chi, \label{eq1}   
\end{equation}
\begin{equation}
D_t\chi = \omega^2_g \partial_x \psi, \label{eq2}
\end{equation} 
where $\nabla^2 \equiv{\partial^2 }/{\partial x^2} + {\partial^2}/{\partial z^2}$, $D_t \equiv \partial_t + {\bf v}\cdot\nabla$, $H$ is the density scale height, $\omega_g$ is the  Brunt-V{\"a}is{\"a}l{\"a} frequency, 
  $\psi(x,z)$ is the velocity potential in which  $z$ represents the vertical direction, and
  $\chi(x,z)$ is the normalized density perturbation. 
  \par 
 Substitution of the expression for the velocity, i.e.,   ${\bf v} = -\partial_z \psi\hat{x} + \partial_x\psi\hat{z}$    into Eqs. \eqref{eq1} and \eqref{eq2}    results in 
\begin{equation}
\begin{split}
\frac{\partial}{\partial t} \nabla^2\psi - \frac{1}{4H^2}\frac{\partial \psi}{\partial t}& = -J(\psi , \nabla^2\psi) - \frac{\partial \chi}{\partial x},\\
\frac{\partial \chi}{\partial t}& = -J(\psi, \chi) + \omega_g^2 \frac{\partial \psi}{\partial x},
\label{eq3}
\end{split}
\end{equation}
where $ J(f, g) = \left({\partial f}/{\partial x}\right)\left({\partial g}/{\partial z}\right) -  \left({\partial g}/{\partial x}\right) \left({\partial f}/{\partial z}\right)$ is the Jacobian. \\
For a class of solutions of Eqs. \eqref{eq3} of the form
\begin{equation}
\begin{split}
\psi&= \left[a(t)\sin(k_0x)+b(t)\cos(k_0x)+ \omega_0\right]z/k_0,\\
\chi& = \left[\alpha(t)\sin(k_0x)+\beta(t)\cos(k_0x)+\gamma(t)\right]z, \label{eq-ansatz}
\end{split}
\end{equation} 
where $k_0$ and $\omega_0$ are constants, Stenflo \cite{stenflo1996} derived the following set of coupled equations for acoustic-gravity waves, given by,
\begin{equation}
\begin{split}
\partial_ta + \Tilde{ \omega_0}b +s_1\beta &= -\nu_1a, \\
\partial_tb - \tilde{ \omega_0}a - s_1\alpha &= -\nu_1b, \\
\partial_t\alpha + \omega_0\beta - s_2b\gamma +\omega^2_g b &= -\nu_2\alpha, \\
\partial_t\beta - \omega_0\alpha +s_2a\gamma -\omega^2_g a &= -\nu_2\beta, \\
\partial_t\gamma +a\beta - \alpha b &= -\nu_3 \gamma.\label{eq-main}
\end{split}
\end{equation}
Here, the  terms containing $\nu_1$ and $\nu_2$ appear when one considers, in addition with the other effects, the  dissipative  terms proportional to $\nabla^4\psi$ and $\nabla^2\psi$  respectively in Eqs. \eqref{eq1} and \eqref{eq2}, and the term proportional to $\nu_3$ corresponds to the damping term.   Also, as in Ref. \citep{stenflo1996}, $\omega_0$ and $\omega_g$ are two control parameters with $\tilde{\omega}_0 = {\omega_0}/\left(1+1/4Hk_0^2\right)=s_1\omega_0$, $s_1= \left(1+1/4H^2k_0^2\right)^{-1}$ and $s_2=1$.
\par
In this paper, we numerically study the dynamical behaviors of Eq. \eqref{eq-main} in absence of the dissipative and damping effects. By means of the linear stability analysis and the Lyapunov exponent spectra,  it is seen that  the nonlinear interaction of acoustic-gravity waves can result into an ordered structure or chaos depending on whether the parameter $s_1$, associated with the density scale height $H$, is below or above its critical value.  
 
\section{Dynamical Properties} \label{sec-dynm-props} In this section, we numerically   study  the dynamical properties of Eqs. \eqref{eq-main}. We  focus mainly on the development of chaos as well as the tendency to form ordered structures in absence of the dissipative effects (i.e., terms proportional to $\nu_1$, $\nu_2$ and $\nu_3$).  Thus, setting $\nu_1=\nu_2=\nu_3=0$ and for convenience, redefining the variables, namely, 
$a=u,~ b = v,~ \alpha = x,~ \beta = y$ and $\gamma = z$, Eq. \eqref{eq-main} can be recast as
%\begin{equation}
%\begin{aligned}
%\frac{du}{dt} &=-{\nu_1}u - \tilde{\omega}_0v -{s_1}y \\
%\frac{dv}{dt} &=-{\nu_1}v +\tilde{\omega}_0u +{s_1}x \\
%\frac{dx}{dt} &=-{\nu_2}x -\omega_0y + vz - {w_g}^2v \\
%\frac{dy}{dt} &=-{\nu_2}y +\omega_0 x - uz + {w_g}^2u \\
%\frac{dz}{dt} &= -uy +xv
%\end{aligned}
%\end{equation} 
\begin{equation}
\begin{split}
\frac{du}{dt} &=-\tilde{\omega}_0v -{s_1}y, \\
\frac{dv}{dt} &=\tilde{\omega}_0u +{s_1}x, \\
\frac{dx}{dt} &=-\omega_0y + vz -  {\omega_g}^2 v, \\
\frac{dy}{dt} &=\omega_0 x - uz +  {\omega_g}^2u, \\
\frac{dz}{dt} &= -uy +xv. \label{eq-reduced}
\end{split}
\end{equation}\\
\subsection{Linear stability analysis}\label{sec-sub-stability-analysis}
In order to perform the stability analysis of the system \eqref{eq-reduced}, we first find its fixed points $\left(u_0,v_0,x_0,y_0,z_0\right)$.  These can be obtained by equating the right-hand sides of Eq. \eqref{eq-reduced} to zero and finding solutions for $u,~v,~x,~y$ and $z$.   
Thus, the fixed points so obtained are  the origin  $O=(0,0,0,0,0)$ and $P=\left(0,0,0,0,\omega_0^2+\omega_g^2\right)$. Next, around the fixed points, we apply the perturbations of the forms: $u'=u-u_0,~v'=v-v_0,~x'=x-x_0,~y'=y-y_0$ and $z'=z-z_0$ to obtain a linearized system of perturbation equations: $d{\bf X}/dt=J{\bf X}$, where ${\bf X}=(u',v',x',y',z')$ and $J$ is the Jacobian matrix. For each fixed point, the eigenvalues $\lambda$ can be obtained from the corresponding eigenvalue problem $J{\bf X}=\lambda{\bf X}$. The stability of the system \eqref{eq-reduced} about the fixed points can then be studied by the nature of these eigenvalues. 
\par
The  Jacobian matrix corresponding to the fixed point $O$ is given by 
\begin{equation}
J_{O} = \begin{bmatrix}
0 &-\tilde{\omega}_0 &0 &-{s_1} &0 \\
\tilde{\omega}_0 &0 &{s_1} &0 &0 \\
0 &- \omega_g^2  &0 &-\omega_0 &0 \\
 \omega_g^2  &0 &\omega_0 &0 &0 \\
0 &0 &0 &0 &0
\end{bmatrix} \label{eq-Jo}
\end{equation}
and the corresponding eigenvalues  of the matrix $J_O$ are given by
 $\lambda=0$ and 
% \begin{equation}
% \begin{split}
% \lambda=&\pm\frac{1}{\sqrt{2}}\left[-(\omega_0^2+\tilde{\omega}_0^2+2s_1\omega_g^2)\right.\\
% &\left.\pm(\omega_0-\tilde{\omega}_0)\sqrt{(\omega_0-\tilde{\omega}_0)^2+4s_1\omega_g^2}{2} \right]^{1/2} 
% \end{split}
% \end{equation}
 %%%%
 \begin{equation}
  \lambda=\pm\frac{1}{\sqrt{2}}\left(-B\pm\sqrt{B^2-4C}\right)^{1/2}, \label{eq-lambda-O} 
  \end{equation}
 %%%
where   $B=\omega_0^2+\tilde{\omega}_0^2+2s_1\omega_g^2$ and $C=\left(s_1\omega_g^2-\omega_0\tilde{\omega}_0\right)^2$.  We note that since $C>0$  and $B> 0$ for  $0<s_1<1$, the values of $\lambda$ in Eq. \eqref{eq-lambda-O} are purely imaginary, i.e.,  $\Re\lambda =0$, implying that the fixed point $O$ corresponds to a stable center.    
\par Next, for the stability of the system \eqref{eq-reduced} around the  fixed point $P$,  we apply the similar perturbations as discussed before, i.e., $u'=u,~v'=v,~x'=x,~y'=y$ but with  $z'=z-\left(\omega_0^2+\omega_g^2\right)$. 
% Therefore the eqn[7] becomes 
% \begin{equation}
%\begin{aligned}
%\frac{du}{dt} &=-\tilde{\omega}_0v -{s_1}y \\
%\frac{dv}{dt} &=\tilde{\omega}_0u +{s_1}x \\
%\frac{dx}{dt} &=-\omega_0y + v(\hat{z}+(\omega_0^2+\omega_g^2)) - {w_g}^2v \\
%\frac{dy}{dt} &=\omega_0 x - u(\hat{z}+(\omega_0^2+\omega_g^2)) + {w_g}^2u \\
%\frac{d\hat{z}}{dt} &= -uy +xv
%\end{aligned}
%\end{equation}\\ 
The corresponding Jacobian matrix $J_P$ and  the corresponding eigenvalues $\lambda$ are, respectively, given by
  \begin{equation}
  J_P = \begin{bmatrix}
0 &-\tilde{\omega}_0 &0 &-{s_1} &0 \\
\tilde{\omega}_0 &0 &{s_1} &0 &0 \\
0 &{\omega_0}^2 &0 &-\omega_0 &0 \\
-{\omega_0}^2 &0 &\omega_0 &0 &0 \\
0 &0 &0 &0 &0
\end{bmatrix}, \label{eq-Jp}
\end{equation}
  $\lambda =0$ and
%$\lambda = \pm\sqrt{\frac{2s_1\omega_0^2-\omega_0^2-\tilde{\omega}_0^2 \pm (\omega_0+\tilde{\omega}_0)\sqrt{\omega_0^2-4\omega_0^2s_1-2\omega_0\tilde{\omega}_0+\tilde{\omega}_0^2}}{2}}$.
\begin{equation}
\lambda = \pm\frac{\omega_0}{\sqrt{2}}\left[-(1-s_1)^2\pm(1+s_1)\sqrt{\left(1-s_1\right)^2-4s_1}\right]^{1/2}, \label{eq-lambda-P}
\end{equation}
 in which we have used the expression $\tilde{\omega}_0=s_1\omega_0$. From Eq. \eqref{eq-lambda-P},   we note that the values of $\lambda$ become purely imaginary for  $0<s_1\lesssim0.17$, and in this case, the the fixed point $P$ corresponds to a stable center. However, for values of $s_1$ in $0.17<s_1<1$, $\lambda$ has complex conjugate values with positive and negative real parts. Thus, it turns out that  the system may be unstable (with at least one $\Re\lambda>0$) around the fixed point $P$ for $0.17<s_1<1$. From the above analysis it follows that the parameter $s_1$, which typically depends on the density scale height $H$ for acoustic-gravity waves, plays a crucial role for the stability and instability of the system \eqref{eq-reduced} about the fixed points $O$ and $P$. In fact, as the the density scale height $H$ increases and so is $s_1$, the system's stability tends to break down, which can lead to the development of chaos as will be shown later.   
 \par
\begin{figure}[h!]
 \begin{center}
         \includegraphics[width=3.2in,height=2in]{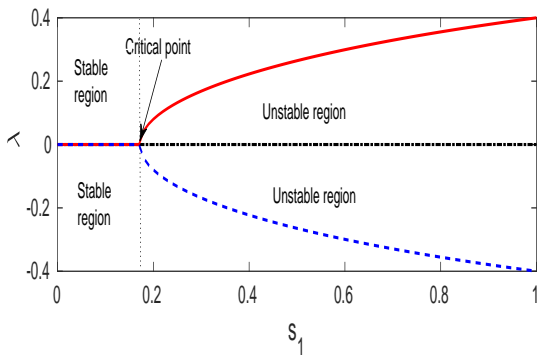}
        \caption{Pitchfork bifurcation diagram showing the stable and unstable regions of the system \eqref{eq-reduced} around the fixed points $O$ and $P$. While the system is stable in the region $0<s_1\lesssim0.17$ where $\lambda=0$ or $\Re\lambda=0$, it exhibits instability in the domain $0.17\lesssim s_1<1$    with  $\Re\lambda>0$. The upper (solid red line) and lower (dashed blue line) branches are corresponding to the $\pm$ sign in the square brackets of the expression for $\lambda$ [Eq. \eqref{eq-lambda-P}]. The other parameter values are $\omega_0 =0.4$, $\omega_g =1.01$,  and $\tilde{\omega}_0 = s_1 \omega_0$. }
        \label{fig:pitchfork} 
        \end{center}      
\end{figure}

 Figure \ref{fig:pitchfork} shows the bifurcation diagram for stable and unstable regions corresponding to the fixed points  $O$ and $P$. We plot $\lambda$, given by Eqs. \eqref{eq-lambda-O} and \eqref{eq-lambda-P}, with respect to the parameter $s_1$  $(0<s_1<1)$. The dash-dotted line represents $\lambda=0$ corresponding to the fixed point $O$. Also, for the fixed point $P$,    $\Re\lambda=0$ in the interval $0<s_1\lesssim0.17$. So, the system is stable around both the fixed points in the domain $0<s_1\lesssim0.17$. However, beyond this domain, i.e., in $0.17<s_1<1$,  the system is shown to be unstable around the fixed point $P$.  In Fig. \ref{fig:pitchfork}, the    upper (solid line) and lower (dashed line) branches   are the plots of $\lambda$ corresponding to the $\pm$ sign in the square brackets in  Eq. \eqref{eq-lambda-P}. 
%In the section of stability analysis we have that for the fixed point $P_1$  system model is stable center  when as $0<s_1<1$  and for $P_2$ a certain value of $s_1$ we have a bifurcation from stable center to unstable character. Hence for the point $P_1$  we get that the jacobian matrix $J_{(0,0,0,0,0)}$ has five eigenvalues  one $0$, and other four eigenvalue are purely imaginary so there is no bifurcation only a stable center . From the figure we have only fixed the $s_1$ parameter value in such a way that we can have figure out the changes around the fixed point $P_2$. In analytically from the deduction  of eigenvalues of the Jacobian matrix  we can get for some values of $s_1$ we get stable center and for some values we get an unstable or chaotic behavior which gives the Pitchfork bifurcation. Numerically we figure out the suitable values in which the eqn[7] perfectly gives the bifurcation. From the figure we get that when $\omega_0 \neq 0$ then the eigenvalue $\lambda$ has positive real parts which implies a unstable behavior and for negative real part it gives the stable center. Again if $\omega_0 = 0$ then all the values of $\lambda = 0$ which gives a stable periodic orbit. 
 \par  
   In the next   subsection  \ref{sec-sub-lyapunov},   we will calculate the Lyapunov exponents spectra to verify the  existence of  chaos with variations of the parameters $s_1$, $\omega_0$ and $\omega_g$.
%\par For the fixed points $P_1$ \& $P_2$ we obtained the eigenvalues of the eqn[7] which imply the Stenflo equation has a strange behavior about the fixed point $P_2$ of course suitable choice of parameters i.e in this choice we can have the chaotic behavior otherwise it gives a stable center in the fig[] we conclude with this two behavior. Again about the fixed point $P_1$  the eqn[7] gives a stable center with any choice of parameter. To solve numerically the eqn[7] with the suitable initial condition and the parameter we have the phase space and lyapunov exponent which provide more precious result of the strange behavior of the eqn[7].
\subsection{Lyapunov exponents} \label{sec-sub-lyapunov} In order to calculate the Lyapunov exponents, we solve the system of equations \eqref{eq-reduced}  with the initial condition $X(0)=\left(u(0),v(0),x(0),y(0),z(0)\right)$. If the system \eqref{eq-reduced} is recast as $\dot{X}=\left(\dot{u}(t),\dot{v}(t),\dot{x}(t),\dot{y}(t),\dot{z}(t)\right)$, its variational form of equation  is given by 
\begin{equation}
\frac{d}{dt}DX(t)=J_L(t)DX(t),\label{eq-variation}
\end{equation}
  where $D\equiv d/dt$,  $DX(0)=I_{5}$ with $I_5$ denoting the identity matrix of order $5$  and $J_L(t)$  the Jacobian matrix evaluated at the initial value $X(0)$,  given  by,
\begin{equation}
J_L(t) = \begin{bmatrix}
0 &-\tilde{\omega}_0 &0 &-{s}_1 &0 \\
\tilde{\omega}_0 &0 &{s}_1 &0 &0 \\
0 &z(t)-\omega_g^2 &0 &-\omega_0 &v(t) \\
\omega_0^2-z(t) &0 &\omega_0 &0 &u(t) \\
-y(t) &x(t) &v(t) &-u(t) &0
\end{bmatrix}.   
\end{equation}
Since $DX(0)$ is non-singular and so is  $DX(t)$,  the solution of Eq. \eqref{eq-variation} is given by
\begin{equation}
\Lambda =\lim_{t\to \infty}\frac{1}{2t}\ln\left[\left(D{X(t)})^TD{X(t)}\right)\right],\label{eq-eigenvalue}
\end{equation}
from which the Lyapunov exponents are obtained as the eigenvalues $\lambda_i,~i=1,...,5$
of the matrix $\Lambda$.  Given a fixed initial condition $X(0)$ of the dynamical system, the change of particle's orbit can be found by the Liouville's formula: $\Delta (t)=\text{tr}(J_L(t))\Delta_t$, where $\Delta_t\equiv\det DX(t)$, $\Delta_0\equiv\det DX(0)=\det I_5=1>0$ and  `tr' denotes the   trace of the matrix $J_L(t)$. Thus, for the system \eqref{eq-reduced} we have
$\det {DX(t)}=\exp(\int_{0}^{t}\text{tr}(J_L(t))dt=1>0$, implying that at least one eigenvalue $\lambda_i$ is positive, and so   a chaotic orbit exists for a certain time period $[0~t]$.
\begin{figure*}[h!]
   \begin{center}
    
        \includegraphics[width=5in,height=2in]{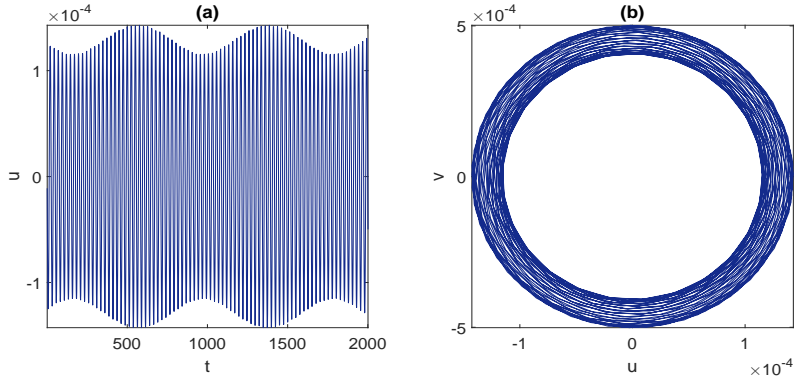}
        \caption{ Stable oscillations: (a) the time series  and (b)  the phase-space diagram   showing that the equilibrium point $O$ corresponds to the stable center. The parameter values are   $\omega_0 = 0.01$, $s_1 = 0.61$, $\omega_g =1.01$ and $\tilde{\omega}_0 = s_1.\omega_0 = 0$. }
        \label{fig:stable-center}
 \end{center}       
\end{figure*}
%%%%%%%%%%%%%%%%%%%%%%%%%%%%%%%%%%%%
%\begin{figure*}[h!]
%    
%        \includegraphics[width=5in,height=3in]{periodictime1.eps}
%        \caption{ Periodic time series for $\omega_0 = 0$, $s_1 = .61$, $\omega_g =1.01$ $\tilde{\omega}_0 = s_1.\omega_0 = 0$ }
%        \label{fig:pireodic}
%      
%\end{figure*}
%
%
%\begin{figure}[h!]
%           \includegraphics[width=4in,height=3in]{bifurcation1.eps}
%        \caption{Phase space diagram of the vectors about the fixed points $P_1$ \& $P_2$ when $\omega_0 = 0 $, given $s_1 = .91$ \& $\omega_g =1.01$ $\tilde{\omega}_0 = s_1.\omega_0 = 0$ }
%        \label{fig:bifur}
%     
%\end{figure}
%%%%%%%%%%%%%%%%%%%%%%%%%%%%%%%%%%%%%%%%%%%%%%%%%%%%
\section{Numerical analysis} 
\begin{figure*}[h!]
   \begin{center}
     
        \includegraphics[width=5in,height=2in]{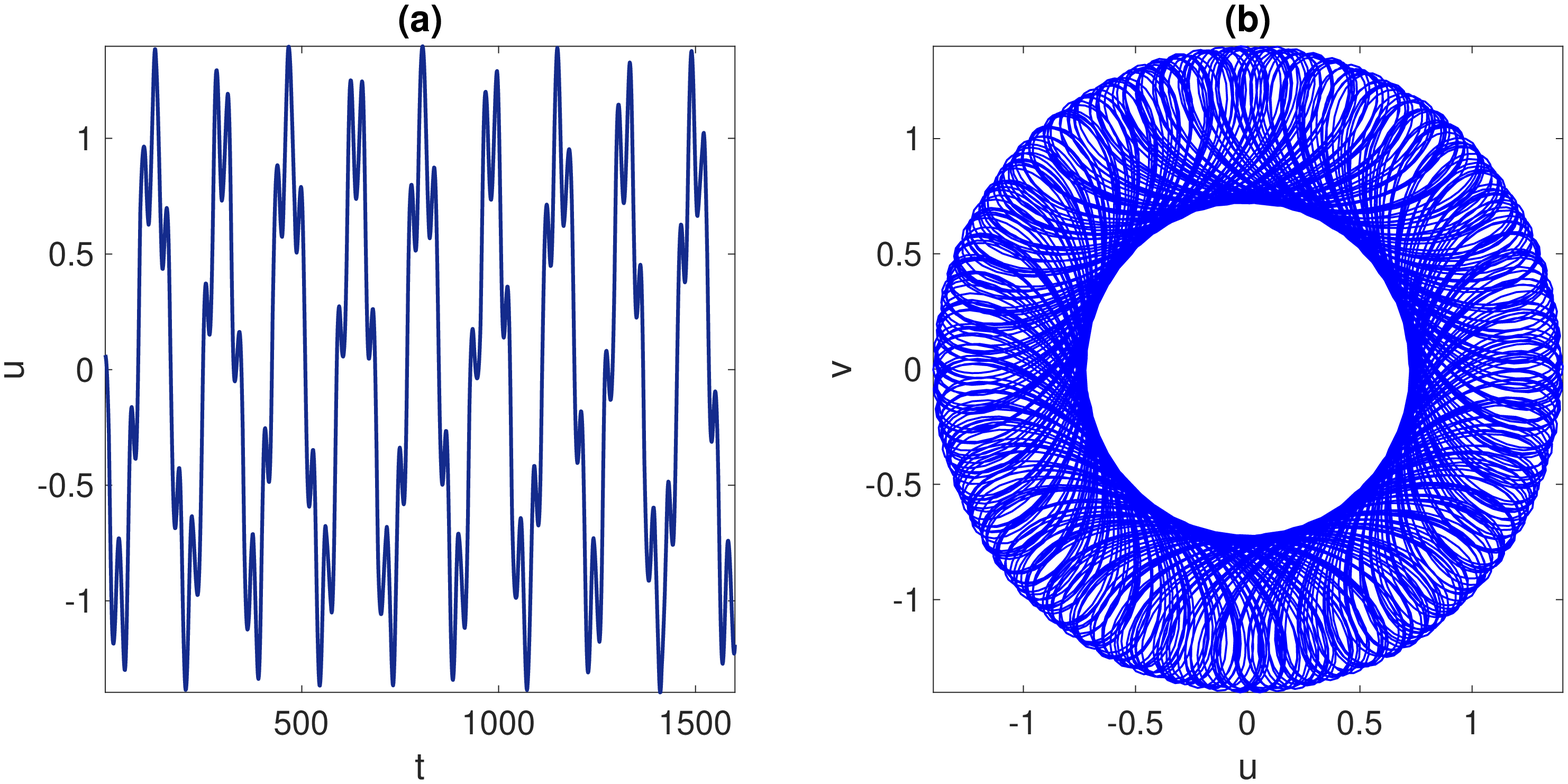}
        \caption{Subplots (a) and (b) are, respectively, the time series   and  the phase space (torus) showing the  quasi-periodicity of the system \eqref{eq-reduced}   with  parameter values $\omega_0 = 1.5$, $s_1 = 0.31$, $\omega_g =1.01$ and $\tilde{\omega}_0 = s_1 \omega_0=0.465$.   }
        \label{fig:lquasi}
 \end{center} 
\end{figure*}

\begin{figure*}[h!]
    \begin{center}
    
        \includegraphics[width=5.2in,height=1.6in]{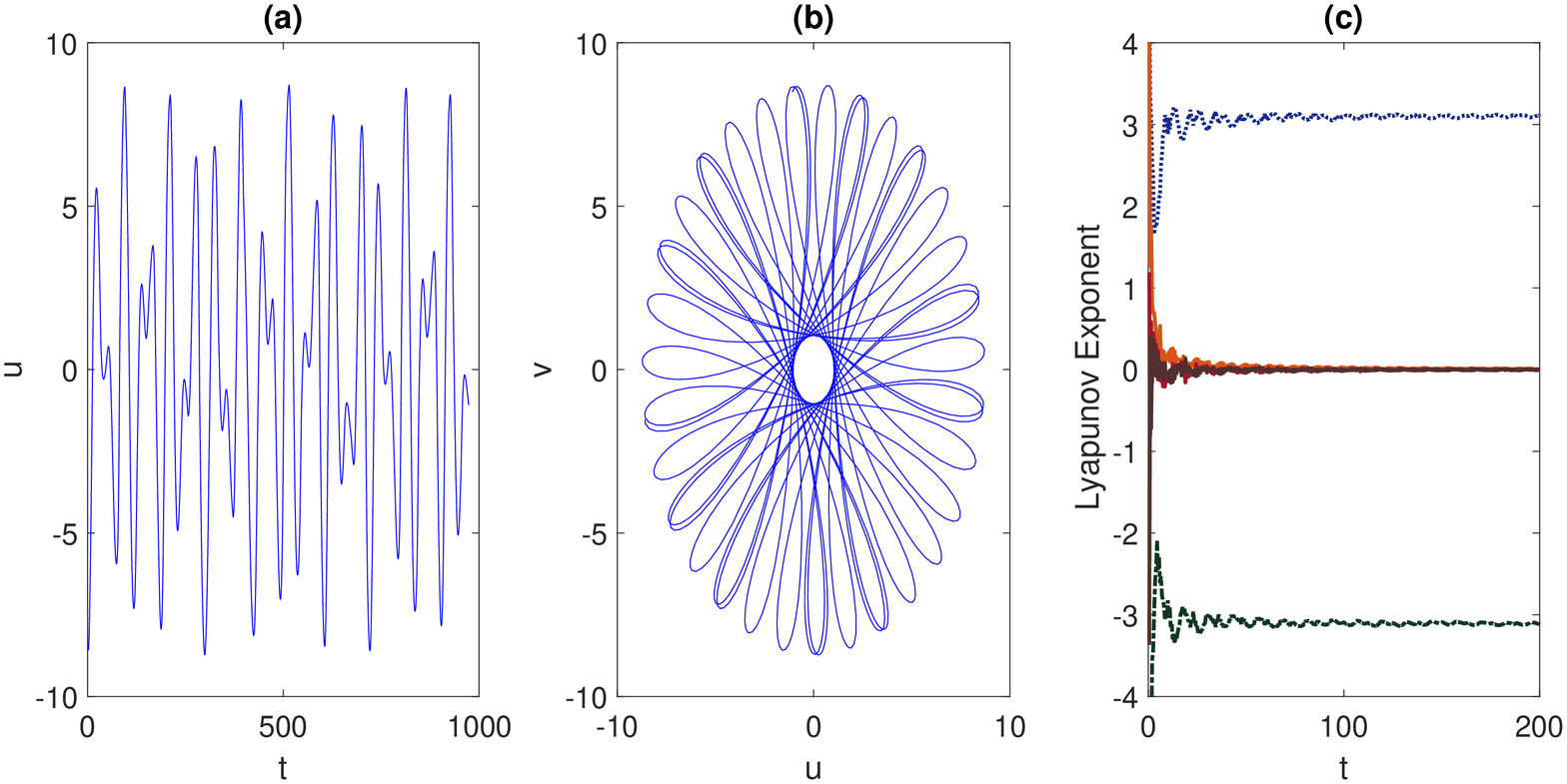}
        \caption{The chaotic time series (a), the chaotic phase space (b) and the Lyapunov exponents (c) are shown with   parameters  $\omega_0 = 1.5$, $s_1 = 0.91$, $\omega_g =1.01$, and $\tilde{\omega}_0 = s_1 \omega_0 =1.365$.  }
        \label{fig:lchaos}
 
    \end{center} 
\end{figure*}

We study the dynamical behaviors of  solutions of the system \eqref{eq-reduced}. To this end, we numerically integrate Eq. \eqref{eq-reduced} by   using the  4-th order Runge-Kutta scheme with a time step $\Delta t = 10^{-3}$.  The results are displayed in Figs. \ref{fig:stable-center} to \ref{fig:lchaos}. We note that for certain ranges of values of the parameters $\omega_0$, $\omega_g$ and $s_1$, the system can exhibit stable solutions together with the  quasi-periodic and chaotic states. We study these behaviors in three different cases as follows.    
 \par 
 \textit{Stable Center:} We note that for $\omega_0=0$, and any values of $\omega_g$ and $s_1$ in $0<s_1<1$, the eigenvalues corresponding to the fixed point $O$ are zero and purely imaginary, while those about the fixed point $P$ are all zero. In this case, the system exhibits stable solutions about the fixed points $O$ and $P$.       The system also possesses a class of stable solutions for $\omega_0>0$,   $\omega_g>0$ and  $0<s_1\lesssim0.17$ ({\it cf.} Sec. \ref{sec-sub-stability-analysis} and the bifurcation diagram in Fig. \ref{fig:pitchfork}). The corresponding time series (a) and the phase space plots (b) are shown in  Fig. \ref{fig:stable-center}.
   \par 
\textit{Quasi-periodicity: }  From the linear stability analysis and the bifurcation diagram (See Fig. \ref{fig:pitchfork}) it is evident that   the system tends to loose its stability for $s_1>0.17$ and any positive values of the frequencies $\omega_0$ and $\omega_g$. In fact, there are two subregions of the parameter  $s_1$:   $0.17<s_1\lesssim s_2$ and $s_2\lesssim s_1<1$. In the former,  the system exhibits quasi-periodicity while in the latter it has chaotic behaviors. However, it is very difficult to find a particular region of $s_1$ in which the quasi-periodicity transits into the chaotic states. Usually, in the quasi-periodic region, we observe a stable torus whereas in the chaotic region, the torus structure breaks down, giving rise to a chaotic  structure.     For a suitable choice of the  initial condition     $(-\omega_0k_1, -\omega_0k, k, k_1,-\omega_0^2-\omega_g^2)$, where  $k = 0.9$ and  $k_1=0.8$ together with the parameters $\omega_0 = 1.5$,    $\omega_g =1.01$ and $s_1 = 0.31$ with $\tilde{\omega}_0 = s_1 \omega_0 =0.465$, Fig. \ref{fig:lquasi} shows that   the torus structure  forms at $s_1=0.31$.   

%%%%%%%%%%%Quasiperiodicity %%%%%%%%%%%%%%%%%%%%%%%%%%%%
%%%%%%%%%%%%%%%%%%%%%%%%%% 
%\begin{figure}[h!]
%    
%        \includegraphics[width=3.5in,height=2.5in]{lquasi.eps}
%        \caption{Lyapunov exponent with respect to time series for quasi-periodic nature with given parameter  $\omega_0 = 1.5$, $s_1 = .31$, $\omega_g =1.01$ $\tilde{\omega}_0 = s_1.\omega_0=.465$ }
%        \label{fig:lquasi}
% 
%\end{figure}
%\begin{figure*}[h!]
%    
%        \includegraphics[width=5in,height=3in]{tquasi.eps}
%        \caption{ Quasi-periodic time series for $\omega_0 = 1.5$, $s_1 = .31$, $\omega_g =1.01$ $\tilde{\omega}_0 = s_1.\omega_0 =.465$ }
%        \label{fig:tquasi}
%  
%    
%\end{figure*}
%~~~~
%\begin{figure}[h!]
%    
%        \includegraphics[width=3.5in,height=3in]{quasiperiodic.eps}
%        \caption{Phase-space for quasi-periodic nature with given parameter  $\omega_0 = 1.5$, $s_1 = .31$, $\omega_g =1.01$ $\tilde{\omega}_0 = s_1.\omega_0=.465$ }
%        \label{fig:quasi}
%  
%\end{figure}
\par 
\textit{Chaotic property:} We note that of the two fixed points $O$ and $P$,  the point $O$   always gives a stable center in   every possible regions of the parameters and the initial conditions. However,  for the other fixed point $P$, we have a stable center  in the region of $0<s_1\lesssim0.17$, while  in the other region $0.17<s_1<1$,  the system exhibits either quasi-periodicity or chaos. For a suitable choice of the initial condition and the parameters, namely, $(-\omega_0k_1, -\omega_0k, k, k_1,-\omega_0^2-\omega_g^2)$ with $k = 5$, $k_1=5.8$, $\omega_0 = 1.5$, $\omega_g =1.01$, $s_1 = 0.91$, and  $\tilde{\omega}_0 = s_1 \omega_0 =1.365$, we show that the system \eqref{eq-reduced}, indeed, exhibits chaos, i.e., the torus which forms at $s_1=0.31$ (see Fig. \ref{fig:lquasi}) breaks down at a higher value of $s_1=0.91$ (Fig. \ref{fig:lchaos}).  The corresponding time series [subplot(a)],    the phase space [subplot(b)] and the Lyapunov exponents [subplot(c)]  are shown in Fig. \ref{fig:lchaos}. Here, the appearance of at least one positive  Lyapunov exponent ensures the existence of chaos.

%%% Chaotic states %%%%%%%%%%%
%%%%%%%%%%%%%%%%%%%%
%\begin{figure}[h!]
%    
%        \includegraphics[width=3.5in,height=2.5in]{lchaos.eps}
%        \caption{Lyapunov exponent with respect to time series for quasi-periodic nature with given parameter  $\omega_0 = 1.5$, $s_1 = .91$, $\omega_g =1.01$ $\tilde{\omega}_0 = s_1.\omega_0 =1.365 $  }
%        \label{fig:lchaos}
%  
%\end{figure}
%\begin{figure*}[h!]
%    
%        \includegraphics[width=4in,height=3in]{tchaos.eps}
%        \caption{ Chaotic time series for $\omega_0 = 1.5$, $s_1 = .91$, $\omega_g =1.01$ $\tilde{\omega}_0 = s_1.\omega_0 =1.365 $  }
%        \label{fig:tchaos}
%    
%\end{figure*}
% 
%\begin{figure}[h!]
%    
%        \includegraphics[width=3.5in,height=3in]{chaos.eps}
%        \caption{Phase-space for chaotic nature with given parameter  $\omega_0 = 1.5$, $s_1 = .91$, $\omega_g =1.01$ $\tilde{\omega}_0 = s_1.\omega_0 =1.365 $  }
%        \label{fig:chaos}
%  
%\end{figure}
%%%%%%%%%%%%%%%%%%
\section{Conclusion}
We have investigated the dynamical properties of five nonlinear coupled Stenflo equations \cite{stenflo1996} that describe the evolution of acoustic-gravity waves in atmospheric disturbances. A linear stability analysis together with the analysis of Lyapunov exponents spectra are carried out for different values of the control parameters.  It is found that   the  parameter  $s_1$, which typically depends on the density scale height of acoustic-gravity waves, plays a vital role for the existence of ordered structures as well as   chaos of the Stenflo equations.  While the system exhibits stable  solutions in the region $0<s_1\lesssim0.17$, it can describe   chaotic behaviors in the other region $0.17<s_1<1$. The present results should be useful for understanding the chaotic properties of the atmospheres of the Earth and other planets.   
%\section{references}
\section*{Acknowledgement} {The authors A. Roy and  A.P. Misra acknowledge  support from UGC-SAP (DRS, Phase III) with Sanction  order No.  F.510/3/DRS-III/2015(SAPI).}


\begin{thebibliography}{100}
\bibitem{stenflo1987} Stenflo, L. 1987. Acoustic solitary vortices.  Physics of  Fluids 30,  3297.
\bibitem{stenflo1991}  Stenflo, L. 1991. Equations describing solitary atmospheric waves.  Physica Scripta 43, 599.
\bibitem{stenflo1995}  Stenflo, L.,  Stepanyants, Yu.A. 1995. Acoustic-gravity modons in the atmosphere.  Annales Geophysicae 13, 973.
\bibitem{stenflo1996} Stenflo, L. 1996. Nonlinear equations for acoustic gravity waves. Physics Letters A 222,  378.
  \bibitem{Jovanovic2002}  Jovanovic, D.,  Stenflo, L.     Shukla, P.K. 2002. Acoustic-gravity nonlinear structures.  Nonlinear Processes in   Geophysics 9, 333. 
 \bibitem{mendonca2015}  Mendonca, J.T.,    Stenflo, L. 2015. Acoustic-gravity waves in the atmosphere:
from Zakharov equations to wave-kinetics. Physica Scripta 90, 055001.
 \bibitem{kaladze2008}  Kaladze, T.D.,  Pokhotelov, O.A.,   Shah, H.A.,   Khan, M.I.,   Stenflo,  L. 2008. Acoustic-gravity waves in the Earth’s ionosphere. Journal of  Atmospheric and  Solar-Terrestrial Physics 70,   1607. 
 \bibitem{park2016}  Park, J.,  Han, B-S,  Lee, H.,  Jeon, Y-L,  Baik, J-J 2016. Stability and periodicity of high-order Lorenz–Stenflo equations.  Physica Scripta 91,    065202.
 \bibitem{banerjee2001}  Banerjee, S.,  Saha, P.,  Roy Chowdhury, A. 2001. Chaotic Scenario in the Stenflo Equations. Physica Scripta 63, 177.
\bibitem{shaikh2008}  Shaikh, D.,  Shukla,  P.K.,    Stenflo, L. 2008. Spectral properties of acoustic gravity wave turbulence. Journal of  Geophysical Research 113,  D06108.



 



%1.	Nonlinear equations for acoustic gravity waves, L.Stenflo, Physics Letters A, Volume 222, Issue 6, 18 November 1996, Pages 378-380\\
%2.	Junho Park, Beom-Soon Han, Hyunho Lee, Ye-Lim Jeon and Jong-Jin Baik, Stability and periodicity of high-order Lorenz–Stenflo equations\\
%3.	L. Stenflo and Yu.A. Stepanyants, Ann. Geophys. 13 (1995) 973.\\
%4.	L. Stenflo, Phys. Fluids 30 (1987) 3297.\\
%5.	Yu.A. Stepanyants, Izv. Atmos. Ocean. Phys. 25 (1989) 729.\\
%6.	E.N. Lorenz. J. Atmos. Sci. 20 (1963) 130.\\
%7.	L. Stenflo, Phys. Ser. 53 (1996) 83.\\
%8.	Zhou C, Lai C H and Yu M Y 1997 J. Math. Phys. 38 5225–39\\
%9.	Shen B W 2014 J. Atmos. Sci. 71 1701–23\\
%10.	Shen B W 2015 Nonlinear Process. Geophys. 22 749–64\\
%11.	Curry J H 1978 Commun. Math. Phys. 60 193–204\\
%12.	Saltzman B 1962 J. Atmos. Sci. 19 329–41\\

\end{thebibliography}
\end{document}